\documentstyle[twocolumn,prl,aps,epsfig]{revtex}
\newcommand{\beq}{\begin{eqnarray}} 
\newcommand{\eeq}{\end{eqnarray}}
\renewcommand{\vec}[1]{{\mathbf{#1}}}
\begin{document} 
\draft 
 
\title 
{Interaction-induced Bose Metal in 2D}
\author{Denis Dalidovich and Philip Phillips}\vspace{.05in}

%
\address
{Loomis Laboratory of Physics\\
University of Illinois at Urbana-Champaign\\
1100 W.Green St., Urbana, IL, 61801-3080}

%
\address{\mbox{ }}
\address{\parbox{14.5cm}{\rm \mbox{ }\mbox{ }
We show here that the regularization of the conductivity
resulting from the bosonic interactions on the `insulating' (quantum disordered)
side of an insulator-superconductor transition in 2D gives
rise to a metal with 
a finite conductivity, $\sigma =(2/\pi) 4 e^2/h$, as
temperature tends to zero. The Bose metal is
stable to weak disorder and hence represents a concrete
example of an interaction-induced metallic phase.  
Phenomenological inclusion of 
dissipation reinstates the anticipated insulating behaviour
in the quantum-disordered regime. Hence,
we conclude that the traditionally-studied 
insulator-superconductor transition, which is driven solely by
quantum fluctuations, corresponds to
a superconductor-metal transition.  The possible relationship
to experiments on superconducting thin films in which a low-temperature metallic
phase has been observed is discussed.    
 }}
\address{\mbox{ }}
\address{\mbox{ }}

\maketitle

\begin{figure}
\begin{center}
\epsfig{file=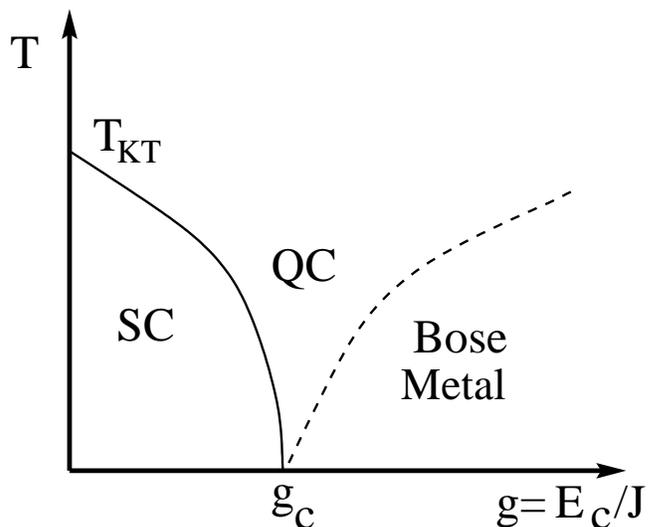, height=7.0cm}
\caption{Phase diagram for the superconductor-Bose metal
transition as a function of temperature
and the quantum fluctuations, $g$.  For
 Josephson-junction arrays, $g$ is given
by the ratio of the charging energy, $E_c$, to the Josephson coupling, $J$.
$T_{\rm KT}$ is the Kosterlitz-Thouless temperature at which phase coherence
obtains.  $g_c$ defines the critical
value of the phase disorder to destroy
the superconducting phase.  QC refers to the quantum critical
regime in which the inverse correlation length is linear in the temperature.}
\label{scc}
\end{center}
\end{figure}

Two ground states are thought to exist for bosons at $T=0$: a superconductor
with long-range phase coherence and an insulator in which the quantum mechanical
phase is disordered.  In this paper, we prove that the phase-disordered
regime is actually not an insulator but rather a metal with a universal
conductivity given by $(2/\pi)4e^2/h$.  This result
arises from a conspiracy: In the quantum-disordered
regime, the population of bosons
is exponentially suppressed; however, so is the scattering
rate between bosons.  But because the conductivity
is a product of the density and the scattering time,
the exponentials cancel, giving rise to a finite
conductivity at $T=0$. 
Any finite amount of dissipation, however, reinstates the
insulator.  Consequently,
the traditional arguments for the insulating
phase of bosons based on quantum fluctuations of the phase must be reconsidered.
Quantum fluctuations alone are insufficient to yield an insulating phase
as illustrated clearly in Fig. (\ref{scc}).  Generically, quantum fluctuations
lead to a Bose metal phase in 2D.  

To establish our result, we rely on the Landau-Ginzburg formalism. While
this approach has had much success\cite{doniach,mpa,fisher,zwerg,chak2,ks,amb,wag2,wag,otterlo,cha} 
in elucidating the critical properties of thin films
and Josephson junction arrays
at the point of transition\cite{fisher,markovic},
the transport properties have proven to be more elusive.
For example, simple physical
considerations\cite{mpa,wen} place the conductivity at the point of transition 
at a universal
value of $\sigma_Q=4e^2/h$, whereas in experiments, $\sigma_Q$ ranges anywhere
from $2\sigma_Q$ to $\sigma_Q/3$\cite{lg,yk,mk}. Likewise, theoretical
calculations based on the Landau-Ginsburg approach have yielded values
ranging from $\pi\sigma_Q/8$\cite{otterlo,cha} to $1.037\sigma_Q$\cite{ds}.
In addition, Damle and Sachdev\cite{ds}
have shown that the frequency
and temperature tending to zero limits do not commute. The
lack of commutativity of these two limits arises from 
the fundamental fact that close to the transition point, the conductivity is a universal 
function of $\hbar\omega/T$\cite{ds}.  Consequently,
$\sigma(\omega\rightarrow 0,T=0)\ne\sigma(\omega=0,T\rightarrow 0)$.
The original theoretical work\cite{mpa,cha} was all based on the former limit which
physically represents the coherent regime.
In such
calculations, the dc-conductivity acquires a singular Drude part on the 
insulating side which can be regularized by the phenomenological inclusion
of dissipation\cite{otterlo,denis}.  
Experimentally, however, it is $\sigma(\omega=0,T\rightarrow 0)$, or equivalently
the hydrodynamic regime,
that is relevant.  In this regime, collisions between quasiparticles
dominates the conductivity which can be obtained from the quantum
kinetic equation.
While such an approach has been used in the quantum critical regime,
no corresponding study has been made in the quantum-disordered 
regime. It is precisely this limit that we study here
using the quantum kinetic approach. 
We show explicitly that the 
quartic interaction between the bosons, whose
dispersion relation is gapped in the phase-disordered
regime,  ultimately gives rise
to a metallic rather than the anticipated insulating phase in the quantum-disordered
regime.  
The Bose metal that 
we find is distinct from the Bose metal of Das and Doniach
\cite{dd} which is purported to obtain once the size of the
Josephson superconducting grains exceeds a critical value.  Further,
our result establishes that the Bose metal is the generic ground
state of bosons lacking phase coherence in the absence of dissipation.

Although our microscopic system is an array of Josephson junctions,
we coarse-grain over the phase associated with each junction and
hence use as our starting point the Landau-Ginzburg action,
\beq\label{gaussian}
F[\psi]&=&\sum_{\vec k,\omega_n}(k^2+\omega_n^2+m^2)
|\psi(\vec k,\omega_n)|^2\nonumber\\
&&+\frac{U}{2N\beta}\sum_{\omega_1,...,\omega_4; \vec k_1,...,\vec k_4}
\delta_{\omega_1+\cdots\omega_4,0} \delta_{\vec k_1+ \cdots \vec k_4,0}\nonumber\\
&&\psi_\nu(\omega_1,\vec k_1)\psi_\nu(\omega_2,\vec k_2)
\psi_\mu(\omega_3,\vec k_3)\psi_\mu(\omega_4,\vec k_4)
\eeq
where $\psi(\vec r,\tau)$ is the complex Bose order parameter
whose expectation value
is proportional to $\langle\exp(i\phi)\rangle$, where $\phi$ is the phase
of a particular junction. The summation in the action
over discrete Matsubara frequencies, $\omega_i=2\pi n_i T$,
and integration over continuous wavevectors, $\vec k$, is assumed. 
The parameter $m^2$ is 
the inverse square of the correlation length. In writing the action in
this fashion, we have already included the one-loop renormalization
arising from the quartic term.  In the quantum-disordered regime,
$m\gg T$ and hence it is the quantum fluctuations that dominate
the divergence of the correlation length.  

Our goal is to calculate the conductivity in the quantum disordered regime
in the collision-dominated limit.  Following the quantum kinetic approach
of Damle and Sachdev\cite{ds}, we introduce the distribution function,
$f(\vec k,t)$ for quasiparticles. We have suppressed the distinction
 between particles
and holes as they have identical distribution functions.  
The quantum kinetic equation for the quasiparticle distribution function 
$f(\vec k,t)$ takes the following form
\beq\label{kin}
\frac{\partial}{\partial t} \delta  f(\vec k ,t) + e^\ast \vec E (t)
 \frac{\partial}{\partial \vec k } n(\epsilon_{\vec k}) = I\{ \delta f \}
\eeq
when it is linearised in the correction $\delta f(\vec k,t)$  to the
equilibrium Bose distribution
 $n(\epsilon_k)=(e^{\epsilon_k/T}-1)^{-1}$ that is induced by
the electric field $\vec E(t)$.  The gapped
dispersion relation for the
bosons is $\epsilon_{\vec k}=\sqrt{k^2+m^2}$.  The resultant linearised 
collision
integral\cite{subir} 
\beq\label{kinetic}
I\{\delta f \} &=& - \frac{1}{N(2\pi )^2}\left[ 
   \int_{\epsilon_{\vec{q}+\vec{k}}>\epsilon_{\vec k}} d^2q
      \frac{1}{\epsilon_{\vec{q}+\vec{k}}\epsilon_{\vec k}} {\rm Im }
       \frac{1}{\Pi (\vec{q}, \epsilon_{\vec{q}+\vec{k}}-\epsilon_{\vec k})}
       \right.\nonumber\\
    &&\left. \left[ \delta 
      f(\vec{k}) ( n(\epsilon_
       {\vec{q}+\vec{k}}-\epsilon_{\vec k}) -n(\epsilon_{\vec q})
      ) \right.\right.\nonumber\\
   &&\left.\left. - \delta f (\vec{q}+\vec{k} ) (1+ n(\epsilon_{\vec k})+
      n(\epsilon_{\vec{q}+\vec{k}}-\epsilon_{\vec k}) ) \right]\right.
      \nonumber\\
  &&\left. + \int_{\epsilon_{\vec{q}+\vec{k}}<
   \epsilon_{\vec k}} d^2 q \frac{1}{\epsilon_{\vec{q}+\vec{k}}
   \epsilon_{\vec k}} {\rm Im}  \frac{1}{\Pi (\vec{q}, \epsilon_{\vec k}-
   \epsilon_{\vec{q}+\vec{k}})}\right. \nonumber\\ 
   &&\left. \times \left[ \delta f (\vec{k}) (1+ n(\epsilon_{\vec q})+
    n(\epsilon_{\vec k}-
      \epsilon_{\vec{q}+\vec{k}}) )\right.\right.\nonumber\\
   &&\left.\left. + \delta f(\vec{q}+\vec{k}) ( 
      n(\epsilon_{\vec k})- n(\epsilon_{\vec k}-\epsilon_{\vec{q}+\vec{k}}) )
 \right]\right.
      \nonumber\\
&&\left.+\int_q d^2q \frac{1}{\epsilon_{\vec q } \epsilon_{\vec k }}
       {\rm Im} \frac{1}{\Pi (\vec q +\vec k ,\epsilon_{\vec q}
+\epsilon_{\vec k } )} \right.\nonumber\\
&&\left. \left[\delta f(\vec k ) (n(\epsilon_{\vec q}) - n(\epsilon_{\vec k}+
       \epsilon_{\vec q} ) )\right.\right.\nonumber\\
&&\left.\left. + \delta f (\vec q) ( n(\epsilon_{\vec k})-
       n(\epsilon_{\vec k}+\epsilon_{\vec q}) ) \right]\right] 
\eeq
is a sum of all incoming and outgoing quasiparticle scattering processes.
A central quantity appearing
in the collision integral is the polarization function
\beq
\label{polariz}
\Pi(\vec q ,i\Omega_m)=T\sum_n \int \frac{d^2p}{(2\pi)^2} 
G_0(\vec p +\vec q, \omega_n +\Omega_m) G_0(\vec p, \omega_n)\nonumber\\
\eeq 
where the bare field propagator $G_0(\vec p, \omega_n)=
(p^2+\omega_n^2+m^2)^{-1}$. As it is the imaginary
part of $\Pi$ that is required in the collision integral, we must perform an 
analytical continuation
$\Omega_n\rightarrow -i\Omega_n -\delta$ with $\delta$ a positive
infinitesimal. 
Unlike Damle and Sachdev\cite{ds}, we have found it convenient to 
work directly in d=2 within a large $N$ expansion. 
Eq. (\ref{kin}) represents the linear integral equation
that must be solved in the generic case to
determine the role of the quartic term on the collision-induced 
conductivity.  

To compute the conductivity, we work in the relaxation-time approximation
in which $I\{\delta f\}\approx -\delta f(\vec k)/\tau_{\vec k}$.  In this 
approximation,
the conductivity emerges as a momentum integration of the form,
\beq
\label{sigma}
\sigma=2\frac{(e^\ast)^2}{\hbar}\int\frac{d^2k}{(2\pi)^2}\frac{k_x^2}
{\epsilon_{\vec k}^2}
\tau_{\vec k}\left(-\frac{\partial n(\epsilon_{\vec k})}
{\partial\epsilon_{\vec k}}\right).
\eeq
We must extract then the relaxation time, $\tau_{\vec k}$ from the collision
integral.  In the quantum-disordered
regime, the statistics of the quasiparticles becomes Boltzmannian because
$m\gg T$. This property results in a suppression of each
subsequent $1/N$ correction by a factor $e^{-m/T}$\cite{chubukov}. 
Having set $N=2$, we obtain then from Eq. (\ref{kinetic}) that
\beq
\label{gentau}
\frac{1}{\tau_{\vec k}} &=& \frac{1}{2(2\pi )^2} \left[   
  \int \frac{d^2 q}{\epsilon_{\vec{q}+\vec{k}}\epsilon_{\vec k}}
  \left( {\rm Im}\frac{1}{\Pi (\vec{q}, \epsilon_{\vec{q}+\vec{k}}-
   \epsilon_{\vec k})} \right) n(\epsilon_{\vec{q}+\vec{k}}-
   \epsilon_{\vec k}) \right.\nonumber\\
  && \left. + \int \frac{d^2 q}{\epsilon_{\vec q}\epsilon_{\vec k}}
    \left( {\rm Im}\frac{1}{\Pi (\vec{q}+\vec{k}, \epsilon_{\vec q}
    +\epsilon_{\vec k})}
    \right) n(\epsilon_{\vec{q}}) \right]
\eeq
to leading accuracy.
Here we took into account the fact that from Eq. (\ref{polariz}),  
 ${\rm Im}\Pi(\vec q,\Omega)$ is an odd function of $\Omega$
and used the identity 
$1+n(\epsilon_{\vec k}-\epsilon_{\vec q +\vec k })=-n(\epsilon_
{\vec q +\vec k }-\epsilon_{\vec k})$.

The essence of our central result is
that to leading order in $T/m$, the inverse relaxation time
$1/\tau_{\vec k}$ is momentum-independent and given by
\beq
\label{relaxtime}
 \frac{1}{\tau}=\pi Te^{-m/T}.
\eeq
Substitution of this expression into Eq. (\ref{sigma}) leads to the mutual 
cancellation of the exponential factors yielding the remarkable result
\beq
\label{mainresult}
 \sigma(T\rightarrow 0)=\frac{2}{\pi} \frac{4e^2}{h}.
\eeq
It is curious to note\cite{qp} that a similar cancellation of exponential
 factors (from the mean free path and the density of states)
arises in the context of the quasiparticle conductivity in a dirty d-wave superconductivity
yielding the identical numerical prefactor $2/\pi$.  

To establish this result, we obtain a workable expression for the inverse
polarization function. In the leading approximation in $T/m$, the real
part of the polarization function can be found from its value
at $T=0$\cite{chubukov}:
\beq\label{real}
\Pi(\vec q,\Omega)=\frac{1}{4\pi\sqrt{q^2-\Omega^2}}{\rm arctan}
\frac{\sqrt{q^2-\Omega^2}}{2m}
\eeq
It is easy to show from the above expression
that when $\sqrt{\Omega^2-q^2}>2m$, the leading  
term of the imaginary part
is temperature independent and equal to
\beq\label{imag2}
 {\rm Im}\Pi(\vec q,\Omega) = -\frac{1}{8\sqrt{\Omega^2-q^2}},
\eeq 
while in all other cases, such a
contribution is absent and the leading temperature dependence comes from
the first term in  
\beq
\label{imag}
{\rm Im}\Pi(\vec q,\Omega)&=&-\frac{1}{16\pi}\int\frac{d^2p}{\epsilon_{\vec p}
\epsilon_{\vec p+\vec q}} \{ |n(\epsilon_{\vec p+\vec q}) - n(\epsilon_
{\vec p})| \nonumber\\
  &&  \delta (|\epsilon_{\vec p+\vec q}-\epsilon_{\vec p}|-\Omega)
  +(n(\epsilon_{\vec p+\vec q})+n(\epsilon_{\vec p})) \nonumber\\
  &&  \delta (\epsilon_{\vec p+\vec q}+\epsilon_{\vec p}-\Omega) \},
\eeq
which is obtained by applying the Poisson summation
formula to Eq. (\ref{polariz})\cite{chubukov}.
The $\delta-$ functions are eliminated upon
an angular integration.  As we are concerned
with the quantum-disordered regime, we focus on the limits
$q, k \leq \sqrt{mT} \ll m$.     
Because in the first
term in Eq. (6), $q>|\epsilon_{\vec{q}+\vec{k}}-\epsilon_{\vec k}|$, 
we must use Eq. (\ref{imag}) to determine the imaginary
part of the inverse polarization function.  Using the fact that in the region of momentum
integration, $|{\rm Re}\Pi(\vec q,\Omega)| \gg |{\rm Im}\Pi(\vec q,\Omega)|$,
we obtain that
\beq
\label{mainpi}
{\rm Im}\frac{1}{\Pi(\vec q, \epsilon_{\vec q+\vec k}-\epsilon_{\vec k})}
&\approx& \frac{8\pi m^2}{q} \sqrt{\frac{\pi T}{2m}} e^{-m/T} \cdot
e^{-(q+(\vec q \vec k )/q)^2 /2mT} \nonumber\\
&& \left( e^{(\epsilon_{\vec q+\vec k}-\epsilon_{\vec k})/T} -1 \right).
\eeq
In the second term of the Eq. (\ref{gentau}) we have that
$[(\epsilon_{\vec q}+\epsilon_{\vec k})^2 -(\vec q +\vec k)^2]>2m$. 
So, it would
be sufficient to use Eq. (\ref{imag2}) for the imaginary part, and
the real part can be obtained from Eq. (\ref{real})
yielding 
\beq
{\rm Im}\frac{1}{\Pi(\vec q +\vec k ,\epsilon_{\vec q}+\epsilon_{\vec k})}   
\approx \frac{16m}{\frac{4}{\pi^2} \ln^2 {\frac{4m}{|\vec q -\vec k|}} +1}. 
\eeq
This suggests that in the limit $m/T \rightarrow \infty$, the contribution
to the relaxation time arising from the second term 
is logarithmically suppressed compared to the contribution from the first
term. Substitution of Eq. (\ref{mainpi}) into Eq. (\ref{gentau}) and
performing the $q-$integration 
we arrive at 
the advertised result, Eq. (\ref{relaxtime}), to leading accuracy in $T/m$.

This result is truly remarkable, because, as one can see, the conductivity
in the quantum disordered regime (in the leading approximation)
depends neither on temperature nor on 
the distance from the transition point.  This means that the dc conductivity
in this model, $\sigma =\sigma_Q g(m/T)$, where $g$ is
a universal function close to 
the transition point, displays a crossover upon lowering $T$ from the 
universal value in the quantum critical regime\cite{ds}, $\sigma_Q$, to 
the smaller value, Eq. (\ref{mainresult}) in the quantum-disordered regime.
Though we focused on the relaxation time approximation,
it is possible to verify {\it a posteriori} 
that the incoming term in the
collision integral provides a contribution subdominant in $T/m$. Indeed, 
seeking the solution of the kinetic equation by means of consecutive
approximations 
$\delta f(\vec k)= \delta f^{(0)}(\vec k)+\delta f^{(1)}(\vec k)+ \cdots$,
where
\beq
\delta f^{(0)}(\vec k) = -e^\ast \tau_{\vec k} \frac{(\vec E \vec k )}
{\epsilon_{\vec k}}
\left( \frac{\partial n}{\partial \epsilon_{\vec k}} \right), 
\eeq   
we obtain after substitution into the kinetic equation, 
that $\delta f^{(1)}(\vec k)$ is proportional to higher powers of
$T/m$ than are the outgoing terms. Consequently, this subdominance justifies
the relaxation time approximation in which only outgoing terms are retained.

The universal value for the conductivity, Eq. (\ref{mainresult}),
was obtained 
for the region near the zero-
temperature IST point, where $m\ll 1$. However, it is not difficult to see
that the above results can be generalized for the case $m \approx O(1)$,
keeping $T\ll 1$. One needs only to make the substitution in the
collision integral
\beq
 \frac{1}{N} {\rm Im} \frac{1}{\Pi (\vec q ,\Omega )} \rightarrow
  {\rm Im} \frac{U}{1+UN\Pi (\vec q ,\Omega )}.
\eeq 
The subsequent steps are identical to those described above and
yield the value of the conductivity
\beq\label{result2}
\sigma(T\rightarrow 0)=\frac{2}{\pi}\frac{U^2}{(4\pi m+U)^2}
\frac{4e^2}{h}.
\eeq
As expected, this value is not universal, and both $m$ and $U$ are the 
functions of parameters of the initial microscopic Hamiltonian.
So, the leveled resistivity, generally speaking, depends on the
distance from the transition point.
In the limit $m\ll 1$,  Eq. (\ref{result2}) reduces to the universal
value (\ref{mainresult}).  Note this result is specific to
2D.  Because the conductivity is not\cite{jose} universal
for the 3D system, the 3D case deserves special attention.

In the approximation that the inverse quasiparticle scattering rate
is small relative to its energy, the general form for the conductivity,
Eq. (\ref{sigma}), can be obtained from the standard
Kubo formula\cite{otterlo,denis}.  With this realization,
the central result, Eq. (\ref{mainresult}), can be 
obtained from Eq. (8) of Ref. (18) by making the substitution
$\eta \rightarrow 1/\tau$ ($\kappa=1$), with $1/\tau$ given
by Eq. (\ref{relaxtime}).  

It is worth exploring how robust our bose metal is
as it has not yet been observed
generically in experiments\cite{van der zant,geerligs,rimberg}.  For 
s-wave pairing, the pair
amplitude survives in the presence of weak disorder\cite{anderson}. 
Consequently, the Bose metal remains intact in this limit. 
As a result, we have constructed a concrete example of an
interaction-induced metal that persists in the limit of weak disorder.
This is important as metallic phases have been observed
in thin films\cite{mk,geerligs,goldman} which should nominally exhibit only insulating
or superconducting phases. For example, Jaeger, et. al.\cite{goldman}
have observed a downturn followed by a leveling of the resistance in
Ga and In thin films with a saturation value ranging between $0.5k\Omega-50k\Omega$.
Our value of $(\pi/2) h/4e^2\approx 10k\Omega$ is certainly within the
 experimental
range over which the Bose metal phase has been observed.
Hence the Bose metal
is a serious candidate to explain these experimental observations. In
addition, the new metallic phase in a dilute 2D electron gas can also be 
explained by
the Bose metal phase if Cooper pairs form at the melting boundary 
of a 2D Wigner
crystal as has been suggested previously\cite{phillips}.  

Experimentally, measurements of the dc-conductivity are always made at finite
frequency.  As the frequency-dependent conductivity has a 
Lorentzian-type peak at $\omega=0$ with a width of order $1/\tau$,
the theoretical constraint on the experimental observation of the Bose metal
is that $\omega\ll 1/\tau$.  For a temperature of $0.1K$, the relaxation
time, Eq. (\ref{relaxtime}) is $10^{10} \exp{(-m/T)}{\rm sec}^{-1}$, where $m\ll T$.
Typical experimental frequencies under which the dc-conductivity is measured
are on the order of several Hz.  Hence, the Bose metal can be observed
provided that $T>0.05 m$.  Typically, $T>0.1m$.
Consequently, there does not appear to be any experimental constraint regarding
the frequency that prohibits the observation of the Bose metal phase.

What about
dissipation?
We have assumed in thus far
that the only source
of quasiparticle relaxation is the quartic term in the 
Ginzburg-Landau action.  Meanwhile, other dissipative mechanisms are
present in real experimental systems exhibiting the IST. 
For example, considering the
dissipation-tuned\cite{rimberg,wag2,wag} IST, Ohmic
dissipation of the form $\eta |\omega_n|$  
is always a relevant perturbation\cite{denis,voelker} arising generically
from coupling to a lattice or an Ohmic resistor.  For d-wave pairing,
static disorder leads to Ohmic dissipation\cite{herbut}.
Provided that $\eta<m$, in the lowest approximation, the
inverse relaxation time $1/\tau$ can be replaced by $1/\tau_{eff}=1/\tau +\eta(k,T)$,
where in the case of Ohmic dissipation $\eta(k,T)=\eta$, and 
our formula for the conductivity
reads in the quantum disordered regime
\beq
 \sigma(T)=2\frac{(e^\ast)^2}{h} \frac{T}{\eta e^{m/T} +\pi T}.
\eeq
We see, that already for very small $\eta$ the conductivity quickly starts 
to decrease exponentially with lowering temperature.  
This result is certainly intriguing and raises
fundamental questions regarding the origin of the insulating state based
on quantum fluctuations.  We propose then that the more correct
phase diagram for insulator-superconductor transitions is Fig. (\ref{scc})
in which it is clearly depicted that quantum fluctuations alone are 
insufficient to drive the
insulating state. 
The finite temperature line starting at $T_{\rm KT}$ and 
terminating at $g_c$ describes the standard Kosterlitz-Thouless transition.
In the absence of dissipation, the $T_{\rm KT}$ line separates a superconductor
and a Bose metal, not a superconductor and an
insulator as had traditionally been thought.  Systematic studies 
are needed in the presence
of weak dissipation are needed 
to see if an insulator replaces the Bose metal phase at low temperatures.

\acknowledgements
We thank the NSF grant No. DMR98-96134 for funding this work
and A. Castro-Neto, S. Sachdev, E. Fradkin, A. Kapitulnik, and A. Yazdani for crucial
discussions.

\end{document}